\documentclass[fleqn,usenatbib]{mnras}

\usepackage{newtxtext,newtxmath}

\usepackage[T1]{fontenc}
\usepackage{ae,aecompl}

\DeclareRobustCommand{\VAN}[3]{#2}
\let\VANthebibliography\thebibliography
\def\thebibliography{\DeclareRobustCommand{\VAN}[3]{##3}\VANthebibliography}

\usepackage{graphicx}	
\usepackage{amsmath}	
\usepackage{newtxtext,newtxmath}
\usepackage{gensymb}    
\usepackage{hyperref}   
\usepackage[table]{xcolor}
\usepackage[flushleft]{threeparttable}

\newcommand{\teff}{$T_{\rm eff}$}
\newcommand{\sigteff}{$\sigma(T_{\rm eff})$}
\newcommand{\logg}{$\log g$}
\newcommand{\siglogg}{$\sigma(\log g)$}
\newcommand{\feh}{$\rm{[Fe/H]}$}
\newcommand{\sigfeh}{$\sigma(\rm{[Fe/H]})$}
\newcommand{\vmic}{$v_{t}$}

\newcommand{\qq}{$\mathrm{q}^{2}$}

\newcommand{\sm}{$\rm{M_{\odot}}$}

\newcommand{\be}{\begin{equation}}
\newcommand{\ee}{\end{equation}}
\newcommand{\ben}{\begin{eqnarray}}
\newcommand{\een}{\end{eqnarray}}
\newcommand{\bfg}{\begin{figure}}
\newcommand{\efg}{\end{figure}}

\newcommand{\kms}{\,km\,s$^{-1}$}

\title[]{Explosive nucleosynthesis of a metal-deficient star as the source of a distinct odd-even effect in the solar twin HIP 11915}

\author[Yana Galarza et al.]
{Jhon Yana Galarza$^{1}$\thanks{E-mail: ramstojh@usp.br},
Jorge Mel\'endez$^{1}$,
Amanda I. Karakas$^{2, 3}$,
Martin Asplund$^{4}$,
\newauthor
Diego Lorenzo-Oliveira$^{1}$
\\
$^{1}$Universidade de S\~ao Paulo, Departamento de Astronomia do IAG/USP, Rua do Mat\~ao 1226, \
     Cidade Universit\'aria, 05508-900 S\~ao Paulo, SP, Brazil \\
$^{2}$Monash Centre for Astrophysics, School of Physics and Astronomy, Monash University Victoria 3800, Australia \\
$^{3}$ARC Centre of Excellence for All Sky Astrophysics in 3 Dimensions (ASTRO 3D) \\
$^{4}$Max Planck Institute for Astrophysics, Karl-Schwarzschild-Str. 1, D-85741 Garching, Germany \\
}

\date{Accepted 24/01/2021. Received 09/01/2021; in original form 30/11/2020}

\pubyear{2021}

\begin{document}
\label{firstpage}
\pagerange{\pageref{firstpage}--\pageref{lastpage}}
\maketitle
\begin{abstract}
The abundance patterns observed in the Sun and in metal-poor stars show a clear odd-even effect. An important question is whether the odd-even effect in solar-metallicity stars is similar to the Sun, or if there are variations that can tell us about different chemical enrichment histories. In this work, we report for the first time observational evidence of a differential odd-even effect in the solar twin HIP 11915, relative to the solar odd-even abundance pattern. The spectra of this star were obtained with high resolving power (140 000) and signal-to-noise ratio ($\sim$420) using the ESPRESSO spectrograph and the VLT telescope. Thanks to the high spectral quality, we obtained extremely precise stellar parameters (\sigteff\ = 2 K, \sigfeh\ = 0.003 dex, and \siglogg\ = 0.008 dex). We determine the chemical abundance of 20 elements ($Z\leq39$) with high precision ($\sim$0.01 dex), which shows a strong pattern of the odd-even effect even after performing Galactic Chemical Evolution corrections. The odd-even effect is reasonably well-reproduced by a core-collapse supernova of 13 \sm\ and metallicity Z = 0.001 diluted into a metal-poor gas of 1 \sm. Our results indicate that HIP 11915 has an odd-even effect slightly different than the Sun, thus confirming a different supernova enrichment history.

\end{abstract}



\section{Introduction}
It is very well known that the composition of even-number elements (C, O, Ne, etc) is higher than the odd-number elements (e.g., N, F, Na) of similar atomic mass. Seminal works observed this effect first in chondritic meteorites \citep{Oddo:doi:10.1002/zaac.19140870118, Harkins:doi:10.1021/ja02250a002}. These authors found a sawtoothed pattern when chemical abundances are plotted versus their respective atomic numbers. Since then, this pattern has became known as the odd-even effect or the Oddo-Harking rule. Later, with the advent of astronomical spectroscopy, it was noticed that the solar abundances are not so different from the Earth and meteorites \citep{Payne:1925PhDT.........1P, Russell:1941Sci....94..375R}. In parallel to these events, new theories based on nuclear shell models began to appear \citep{Mayer:PhysRev.74.235, Mayer:PhysRev.75.1969, Haxel:PhysRev.75.1766.2}. However, it was not until the 1950s that the basis for the development of theories of stellar \textit{nucleosynthesis} \citep[e.g.,][]{Burbidge:1957RvMP...29..547B, Cameron:1959PCE.....3..199C, Cameron:1959ApJ...130..429C} were established thanks to the cosmic abundance distribution of elements compiled by \citet{Suess:1956RvMP...28...53S}. Inspired by these works, more precise abundance compilations were put together, taking into account not only elements from meteorites, but also from the solar photosphere \citep[e.g.,][]{Cameron:1973SSRv...15..121C, Anders:1989GeCoA..53..197A, Grevesse:1998SSRv...85..161G, Lodders:2003ApJ...591.1220L, Asplund:2005ASPC..336...25A, Lodders:2009LanB...4B..712L, Asplund:2009ARA&A..47..481A}. All these results only confirm that the odd-even effect is real in the Sun, thereby suggesting that the proto-cloud of the Sun has this nucleosynthetic signature.

\begin{table*}
	\caption{Comparison of high precision spectroscopic stellar parameters for HIP 11915.}
	\label{tab:fundamental parameters}
	\centering
	\begin{tabular}{cccccccc} 
		\hline
		\hline
		\teff         & \logg             & \feh               & \vmic            & Age             & Mass              & \logg$^{\dagger}$            & Reference    \\
		(K)           & (dex)             & (dex)              &  (\kms)   & (Gyr)            & \sm               & (dex)               &           \\
		\hline 
		5773 $\pm$ 2  & 4.470 $\pm$ 0.008 & $-$0.057 $\pm$ 0.003 & 1.02 $\pm$ 0.01  & 3.87 $\pm$ 0.39$^{\star}$ & 0.991 $\pm$ 0.003 & 4.483 $\pm$ 0.022 & This work \\
		5769 $\pm$ 4  & 4.480 $\pm$ 0.011 & $-$0.067 $\pm$ 0.004 & 0.99 $\pm$ 0.01  & 3.40 $\pm$ 0.60$^{\star}$ & 0.993 $\pm$ 0.007 & 4.482 $\pm$ 0.023 & \citet{Spina:2018MNRAS.474.2580S} \\
		5760 $\pm$ 4  & 4.460 $\pm$ 0.010 & $-$0.059 $\pm$ 0.004 & 0.97 $\pm$ 0.01  & 4.00 $\pm$ 0.60$^{\ast}$  & 0.993 $\pm$ 0.005 & 4.480 $\pm$ 0.023 & \citet{Ramirez:2014AA...572A..48R} \\
		\hline
		\hline
	\end{tabular}
	\begin{tablenotes}
	\item \textbf{Notes.} $^{(\star)}$ Isochronal ages estimated using \logg\ and parallaxes as priors (\logg\ \& plx). $^{(\ast)}$ Isochronal age determined using only \logg\ as input parameter. $^{(\dagger)}$ Trigonometric \logg\ obtained using \textit{Gaia} EDR3 parallax and bolometric corrections of \citet{Melendez:2006ApJ...641L.133M}.
	\end{tablenotes}
\end{table*} 

There are two types of supernova explosions, Ia and II. The first is a result from the explosion of a binary system including a white dwarf,  while the second is from a massive star with mass above $\sim$10 \sm\ and undergoing core-collapse at the end of their evolution \citep{Nomoto:2013ARA&A..51..457N}. During the supernova explosion, enormous energies are liberated and newly synthesized elements are ejected to the interstellar medium, enriching the chemical abundances of the Galaxy. These energetic events established the basis for our understanding of Galactic Chemical Evolution (GCE). Stars with masses between 10-13 \sm\ explode as faint supernova \citep{Nomoto:2013ARA&A..51..457N}, and those with masses between 13-25 \sm\ explode as \emph{normal supernovae} with explosion energy $E_{51} = E/10^{51}$ ergs = 1 \citep{ Blinnikov:2000ApJ...532.1132B}. The odd-even pattern from early supernovae can be seen in metal-poor stars, as shown by different theoretical and observational studies \citep[e.g.,][]{Umeda:2003Natur.422..871U, Iwamoto:2005Sci...309..451I, Tominaga:2007ApJ...660..516T, Heger:2010ApJ...724..341H, Nomoto:2013ARA&A..51..457N, Siqueira:2015A&A...584A..86S, Placco:2016ApJ...833...21P, Frebel:2019ApJ...871..146F}.

Although the odd-even pattern in the Sun has been known for a long time, to our knowledge no variations have been reported for solar-metallicity stars, relative to the Sun's odd-even pattern. In this work we report for the first time observational evidence of a distinct odd-even effect in a solar-metallicity star, thus providing important clues to a distinct chemical enrichment history to our Sun.

\section{Observations and data reductions}
The spectra of HIP 11915 were obtained using the ESPRESSO (Echelle SPectrograph for Rocky Exoplanets and Stable Spectroscopic Observations) spectrograph \citep{Pepe:2014AN....335....8P} and the Very Large Telescope (8.2 m telescope) at the Paranal Observatory between 2018 and 2019, under the ESO (European Southern Observatory) program ID 0102.C-0523. The spectra of the Sun (reflected light from the Vesta asteroid) were downloaded from the ESO Science Archive Facility under the program ID 1102.A-0852. The instrument was configured in its High Resolution 1-UT (HR) mode to reach a high resolving power ($R = \lambda / \Delta \lambda = 140\ 000$). The ESPRESSO spectra were reduced by a pipeline using the EsoReflex environment \citep{Freudling:2013A&A...559A..96F}. The resulting spectra cover the entire visible wavelength ranging from 3 800 to 7 880 \AA.

The reduced ESPRESSO spectra were normalized using our semi-automatic PyRAF\footnote{PyRAF is a product of the Space Telescope Science Institute, which is operated by AURA for NASA.} scripts that divide each spectrum in several regions to normalize them with order polynomials ranging from 1 to 7 using the task \texttt{continuum} of IRAF and always taking as reference the normalized continuum of the Sun. The script finds a solution when the ratio between the continuum of the Sun and the star is approximately one. Finally, the script combines all the normalized spectra using the \texttt{scombine} task in order to achieve the highest possible signal-to-noise ratio (SNR). The resulting SNR for the Sun and HIP 11915 is $\sim$320 and $\sim$420, respectively.

\section{Spectroscopic analysis}
\subsection{Fundamental Parameters}
The reduced equivalent widths ($EW$s) were measured with our python script (for more details see Yana Galarza et al. 2021, submitted). In summary, the script uses the line list from \citet{Melendez:2014ApJ...791...14M} to plot the spectra of the Sun and HIP 11915 in windows of 6 \AA\ with the line of interest located at the center. Then, the $EW$ is measured manually, on a line-by-line basis, through Gaussian fits to the line profile using the Kapteyn \texttt{kmpfit} Package \citep{KapteynPackage}. This method is based on the differential technique between the star and the Sun and allow us to achieve a high precision \citep[e.g.][]{Melendez:2012A&A...543A..29M, Spina:2018MNRAS.474.2580S, Yana:2019MNRAS.490L..86Y}. The script creates an output file containing information about the local continuum, limits of the Gaussian fits, $\chi^{2}$ test, excitation potential (eV), oscillator strength, and laboratory $\log (gf)$ values, as well as the hyperfine structure information when necessary. 

\begin{figure*}
 \includegraphics[scale=0.5]{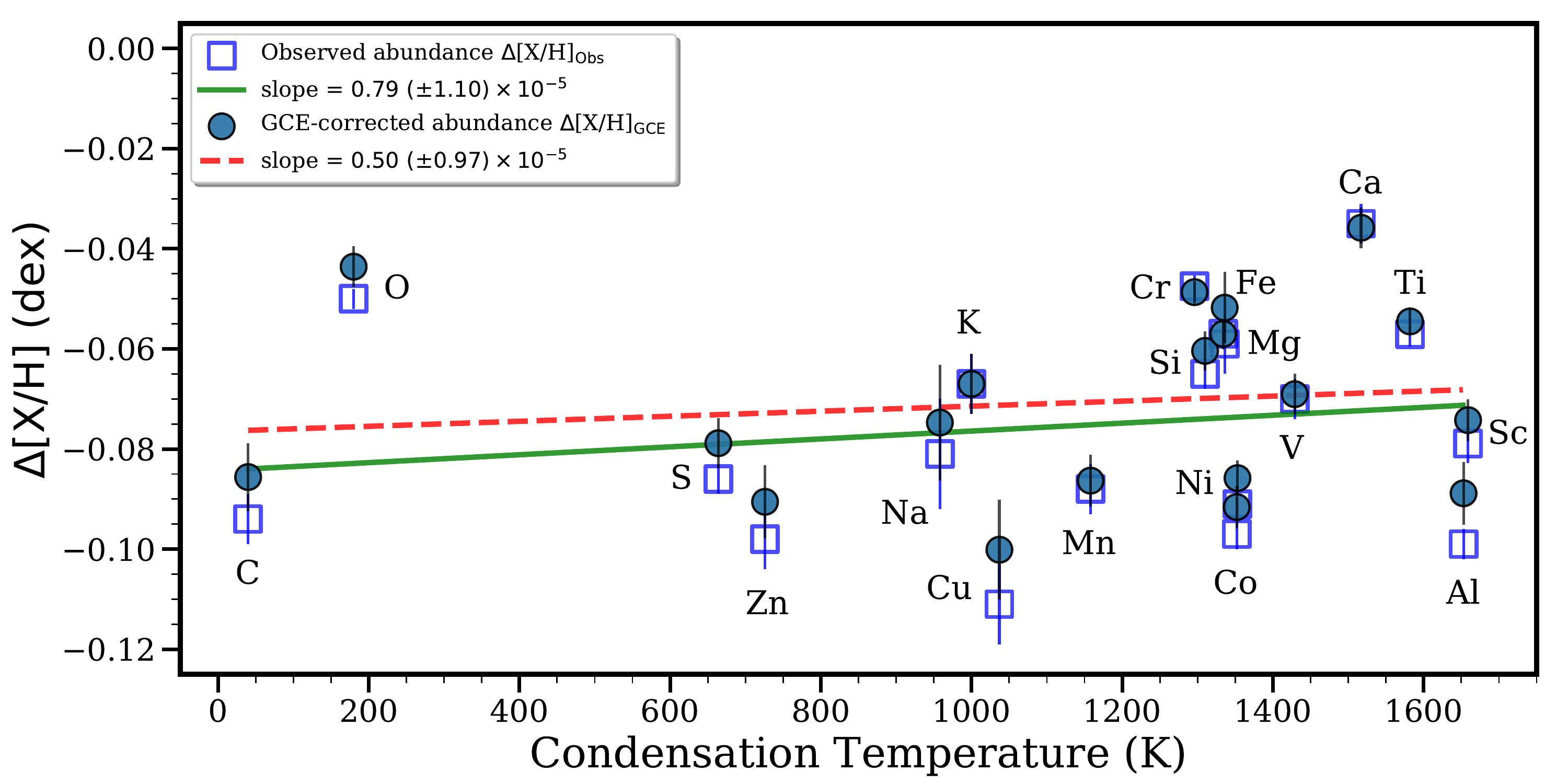}
 \centering
 \caption{Observed (open squares) and GCE-corrected (filled circles) differential abundances relative to the Sun versus condensation temperature, and with their linear fits represented as green solid and red dashed lines, respectively.}
 \label{fig:tc_ab_espresso}
\end{figure*}


We estimated the spectroscopic stellar parameters (effective temperature \teff, surface gravity \logg, metallicity \feh, and microturbulence velocity \vmic) using the automatic qoyllur-quipu python code (hereafter \qq; \citet{Ramirez:2014AA...572A..48R}). The code determines the abundance of 117 spectral iron lines \citep[Fe I and Fe II taken from][]{Melendez:2014ApJ...791...14M} using the 2019 version of the local thermodynamic equilibrium (LTE) code MOOG \citep{Sneden:1973PhDT.......180S} and the Kurucz ODFNEW model atmospheres \citep{Castelli:2003IAUS..210P.A20C}. Then, the stellar parameters are calculated through the spectroscopic equilibrium technique, which consists in the strict fulfillment of four criteria: 1) non-dependence of the differential iron abundance (HIP 1195 - Sun) with the excitation potential, i.e., the slope of them should be zero; 2) non-dependence of the differential iron abundance with the reduced equivalent width (slope consistent with zero); 3) the differential abundance of Fe I and Fe II should be equal; 4) the input metallicity of the model should have the same value as the derived iron abundance. The errors for the stellar parameters are estimated following the prescription given in \citet{Epstein:2010ApJ...709..447E} and \citet{Bensby:2014A&A...562A..71B}), that is from the propagation of the error associated with the fulfillment of the above conditions. Our results are summarized in Table \ref{tab:fundamental parameters} and compared with those from the literature \citep{Ramirez:2014AA...572A..48R, Spina:2018MNRAS.474.2580S}. There is an excellent agreement between them, thus validating the results of previous high-precision works that used spectrographs of lower $R$ than ESPRESSO (HARPS and MIKE with $R$ = 115 000 and 83/65 000, respectively).

The isochronal fitting technique has demonstrated to be a powerful tool to determine ages and masses  \citep[e.g.,][]{Lachaume:1999A&A...348..897L, Takeda:2007A&A...468..663T}. \cite{Ramirez:2013ApJ...764...78R} in their \qq\ code replaced the $M_{V}$ by precise spectroscopic \logg, that improved the results relative to  the use of uncertain parallaxes from Hipparcos. Later, \citet{Spina:2018MNRAS.474.2580S} included \textit{Gaia} parallaxes and effects of the $\alpha$-enhancements into the \qq\ calculations (hereafter \logg\ \& plx). Both methods achieve a high internal precision with uncertainties ranging from $\sim$1-2 Gyrs. We estimated the age and mass of HIP 11915 using the \qq\ code and the Yonsei-Yale isochrone set \citep{Yi:2001ApJS..136..417Y, Demarque:2004ApJS..155..667D}, following the prescription given by \citet{Spina:2018MNRAS.474.2580S}, employing both parallaxes and spectroscopic \logg. The parallax value adopted in our calculation was taken from \textit{Gaia} EDR3 \citep{GAIADR3:2020arXiv201201533G}. In Table \ref{tab:fundamental parameters} is shown the good agreement between our age and mass results with those from \citet{Spina:2018MNRAS.474.2580S} and \citet{Ramirez:2014AA...572A..48R}. As a consequence of our extremely precise spectroscopic stellar parameters, the internal precision of the age and mass are the lowest reported in the literature. However, special attention should be paid in this point because these reflect only the precision of the differential method employed. As the spectroscopic \logg\ is fundamental to determine isochronal ages, we estimated the trigonometric \logg\ using parallax from \textit{Gaia} EDR3 and adopting the bolometric corrections of \citet{Melendez:2006ApJ...641L.133M}. The spectroscopic and trigonometric \logg\ are in very good agreement (see Table \ref{tab:fundamental parameters}), thereby ruling out any problem in the $EW$ measurements, or the spectroscopic method. Similar trigonometric \logg\ values are calculated for \citet{Spina:2018MNRAS.474.2580S} and \citet{Ramirez:2014AA...572A..48R} using their own spectroscopic stellar parameters.

\begin{table}
	\caption{Observed abundances $\Delta$[X/H]$_{\rm{Obs}}$ and GCE-corrected abundances $\Delta$[X/H]$_{\rm{GCE}}$ of HIP 11915 relative to the Sun and their corresponding errors.}
	\label{tab:odd-even expresso}
	\begin{tabular}{ccccc} 
		\hline
		\hline
		element   & Z  &   $\Delta$[X/H]$_{\rm{Obs}}$  & $\Delta$[X/H]$_{\rm{GCE}}$ & [X/H]$_{\rm{SN}}$  \\
		\hline                                                                    
		  C       & 6  &     $-0.094 \pm 0.005$       &     $-0.086 \pm 0.007$      &    $-0.063$   \\
		  O       & 8  &     $-0.050 \pm 0.002$       &     $-0.044 \pm 0.004$      &    $-0.036$   \\
		  Na      & 11 &     $-0.081 \pm 0.011$       &     $-0.075 \pm 0.012$      &    $-0.080$   \\
          Mg      & 12 &     $-0.059 \pm 0.006$       &     $-0.052 \pm 0.007$      &    $-0.033$   \\
          Al      & 13 &     $-0.099 \pm 0.003$       &     $-0.089 \pm 0.006$      &    $-0.047$   \\
		\hline
		\hline
	\end{tabular}
	\begin{tablenotes}
	\item This table is available in its entirety in machine readable format at the CDS.
	\end{tablenotes}
\end{table} 

\begin{figure*}
 \includegraphics[scale=0.50]{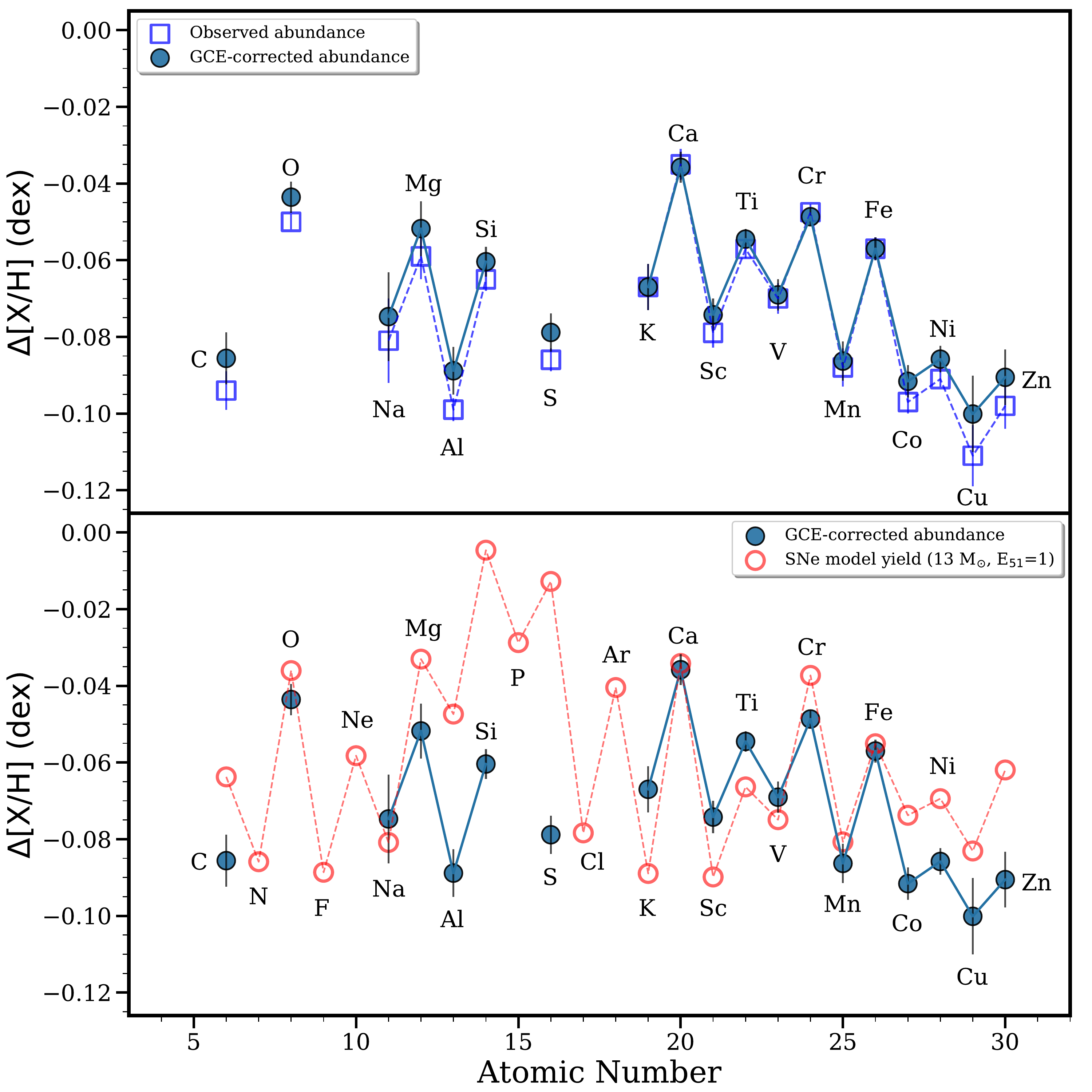}
 \centering
 \caption{\textbf{Upper panel:} Comparison of the chemical-abundance pattern before (blue open squares) and after (blue filled circles) performing GCE corrections. \textbf{Bottom panel:} Comparison between the SN model of 13 \sm\ (red open circles) and the GCE-corrected differential abundances (blue filled circles). The solid and dashed lines in both panels are only to highlight the odd-even effect in HIP 11915.}
 \label{fig:odd_even_espresso}
\end{figure*} 

As a sanity test, the age of HIP 11915 was also estimated using other methods. We determined an age of 3.42 ($\pm$0.43) Gyr and 3.32 ($\pm$0.41) Gyr employing the [Y/Mg]-age correlation independently established by \citet{Spina:2018MNRAS.474.2580S} and \citet{Nissen:2016A&A...593A..65N}, respectively. In addition, according to the activity-age relation of \citet{Diego:2018A&A...619A..73L}, the chromospheric age is 3.6$_{-0.4}^{+0.5}$ Gyr. All these independent age results are in good agreement with our isochronal age, thus confirming the high precision achieved in this work.

\subsection{Abundance pattern}
We measured the extremely high-precision chemical abundance of 19 elements with atomic number lower than 30, plus Y (Z = 39) for age determination. We configured the \qq\ code to perform hyperfine corrections for the elements V, Mn, Co, Cu, and Y following the data given in \cite{Melendez:2014ApJ...791...14M, Asplund:2009ARA&A..47..481A, McWilliam:1998AJ....115.1640M, Cohen:2003ApJ...588.1082C}. The NLTE corrections were only carried out for the O I triplet using the grid of \cite{Ramirez:2007A&A...465..271R}. Despite that there are more updated NLTE corrections \citep[e.g.][]{Amarsi:2016MNRAS.455.3735A, Amarsi:2019A&A...630A.104A}, these are negligible for our purpose since our analysis is strictly differential to the Sun \citep[for more details see][]{Yana_Galarza:2016A&A...589A..17Y}. On the other hand, the interstellar medium is constantly enriched by different sources that produce new elements through stellar nucleosynthesis, and this enrichment mechanism is observed in the chemical abundances of solar twins as a function of time \citep[e.g., ][]{Nissen:2015A&A...579A..52N, Nissen:2016A&A...593A..65N, Spina:2016A&A...593A.125S, Spina:2018MNRAS.474.2580S, Bedell:2018ApJ...865...68B}. Therefore, we performed a GCE correction using the [X/Fe]-age correlation of \citet{Bedell:2018ApJ...865...68B} by adding GCE trends to the chemical abundance of HIP 11915 following the prescription given in \citet{Spina:2016A&A...585A.152S} and \citet{Yana_Galarza:2016A&A...589A..17Y}, so that the abundances of HIP 11915 are corrected to the same age as the Sun. 

In Figure \ref{fig:tc_ab_espresso} we show the classical differential chemical abundance ($\Delta$[X/H]) versus condensation temperature \citep{Lodders:2003ApJ...591.1220L} plot, where the open squares and filled circles are the observed ($\Delta$[X/H]$_{\rm{Obs}}$) and GCE-corrected abundances ($\Delta$[X/H]$_{\rm{GCE}}$), respectively. We can see that the GCE corrections are very small because of the similarity in age (and \feh) between HIP 11915 and the Sun. The lines represent linear fits to the data (using the Kapteyn \texttt{kmpfit} Package), whose slopes are both flat within the errors, meaning that both abundance patterns are similar to the Sun, as reported also in the preliminary analysis of \citet{Bedell:2015A&A...581A..34B}. This is remarkable, since HIP 11915 hosts a Jupiter twin \citep{Bedell:2015A&A...581A..34B}, being thus potentially a \emph{solar system twin}. Intriguingly, there is quite a high scatter ($\sim$0.020 dex) in the observed abundance, which is significantly larger than the average error bar (0.005 dex), and this still prevails in the GCE-corrected abundances ($\sim$0.018 dex). 

The upper panel of Figure \ref{fig:odd_even_espresso} shows the $\Delta$[X/H] versus their atomic numbers. Despite that the estimated abundance is relative to the Sun, our precise abundances allow to clearly see a stronger signature of the odd-even effect, and the pattern still persists in the GCE-corrected abundances (blue filled circles), thus suggesting that this is a chemical signature of a different source. It is important to highlight that we could not determine abundances for N, F, Ne, P, Cl and Ar from our spectra. Only N is marginally available at 7468.31 \AA, however the quality of the spectrum is not adequate for a high precision analysis. A tentative analysis suggests that is more depleted than carbon (N = $-$0.24 $\pm$ 0.05 dex), therefore also following the predicted odd-even pattern. Atomic diffusion could introduce small variations in the abundance pattern \citep[see][]{Dotter:2017ApJ...840...99D}, but the differential effect should be minor for stars of similar ages. Furthermore, the GCE corrections follow the observational trends with age, probably including also the effect of atomic diffusion.

We tried different supernova yields to match our observations. The best model \citep{Tominaga:2007ApJ...660..516T, Nomoto:2013ARA&A..51..457N} that reproduces the odd-even pattern of HIP 11915 is a core-collapse SN with a progenitor mass of 13 \sm\ and metallicity Z = 0.001 diluted into a protocloud of metal-poor gas \citep[80\% of solar,][]{Asplund:2009ARA&A..47..481A} of 1 \sm. We estimated a dilution of 1.5\% mass of SN material to match the iron abundance of HIP 11915. The solar metallicity yields adopted for the model are taken from \citet{Nomoto:2013ARA&A..51..457N} (online table, therein). As can be seen in the bottom panel of Figure \ref{fig:odd_even_espresso} (red open circles), the alpha elements (O, Mg, Ca, Ti), the iron-peaks elements (Sc, V, Cr, Mn, ,Fe) and Na are the best reproduced by the model. More massive SNe model ($>$13 \sm) overproduce the alpha elements.

Despite that some elements such as Al, Si, S and K show an offset, they are qualitatively in agreement with the observations. Our results hint that the observed odd-even effect is a direct nucleosynthetic signature of a particular SN, thereby suggesting that HIP 11915 experienced a distinct chemical enrichment than the Sun. All our differential abundances and those estimated from the SN model ([X/H]$_{\rm{SN}}$) are summarized in Table \ref{tab:odd-even expresso}. 

\section{Summary and Conclusions}
In this work we determined the stellar parameters of the solar twin HIP 11915 with unprecedented precision (see Table \ref{tab:fundamental parameters}) using the ESPRESSO spectrograph at the VLT. The high resolving power ($R$ = 140 000), high signal-to-noise ratio (SNR = 420), and the differential technique relative to the Sun allow us to achieve an extremely high internal precision (\sigteff\ = 2 K, \sigfeh\ = 0.003 dex, and \siglogg\ = 0.008 dex). A precise isochronal age was estimated using isochrones compared to our stellar parameters (including \logg) and the \textit{Gaia} EDR3 parallax. As a sanity check, we also estimated the age using the [Y/Mg]- and the activity-age correlation, which are in good agreement with our resulting age. We determined the high-precision differential chemical abundances of 19 light elements with atomic number up to 30, and the heavy element Y for the age determination. We also carried out GCE corrections, but the odd-even effect persists in the differential abundance of HIP 11915 (blue filled circles in Figure \ref{fig:odd_even_espresso}), thus suggesting a stronger odd-even effect than in the Sun. The distinct odd-even pattern, with a peak-to-peak amplitude of only $\sim$0.02 dex, is revealed only thanks to the high precision of our work. A mechanisms that explain well the GCE-corrected differential abundances is a core-collapse SN of 13 \sm\ and metallicity Z = 0.001 diluted into a 1 \sm\ of metal-poor gas cloud (see open red circles in the bottom panel of Figure \ref{fig:odd_even_espresso}). Albeit a few elements show offsets, overall they follow qualitatively the stronger odd-even pattern. 

The main motivation of large spectroscopy surveys (e.g., \textit{APOGEE}, \textit{Gaia} and \textit{GALAH}) is chemical tagging, aiming to identify coeval stellar groups to reconstruct the Milky Way’s formation. However, as showed in this work, only an analysis with a spectra of extremely high quality can discern the very subtle abundance pattern due to differences in nucleosynthetic enrichment and thus better identify coeval stars. The comparison between the SN model and the abundance pattern of HIP 11915 provides new insights to understand the fine structure in stellar abundances coming from a particular SNe nucleosynthesis history, as well as to improve SNe yields. In addition, \citet{Mackereth:2018PASP..130k4501M} estimate that HIP 11915's Galactic orbit has a maximum vertical excursion of 0.2643 $\pm$ 0.0027 kpc, an eccentricity of 0.1867 $\pm$ 0.0006, and perigalacticon and apogalacticon radii of 5.8036 $\pm$ 0.0063 and 8.4695 $\pm$ 0.0025 kpc, respectively. These results indicate that HIP 11915 belongs to the thin disk (considering the thin disk scale height as 0.36 kpc, \citet{Sanders:2015MNRAS.449.3479S}), but with a significantly higher eccentricity than the Sun ($\sim$0.06 - 0.10, \citet{Bovy:2012ApJ...759..131B}). This provides evidence that the natal cloud of HIP 11915 may have experienced a somewhat different chemical enrichment history.  Despite the somewhat different kinematics, we consider that HIP 11915 is an excellent proxy of the Sun, with similar spectroscopic fundamental parameters, akin age ($\sim$4 Gyr), mass ($\sim$0.99 \sm), and activity index \citep[$\langle S_{\rm{MW}}\rangle$ = 0.187,][]{Diego:2018A&A...619A..73L}. Considering that a Jupiter twin has been detected in this solar twin \citep{Bedell:2015A&A...581A..34B} and that we find in this work an overall abundance pattern versus dust condensation temperature similar to the Sun, being thus also depleted in rocky-forming elements, we consider HIP 11915 an excellent system for planet searches of Earth analogs.

\section*{Acknowledgements}
\textit{Research funding agencies:} J.Y.G. acknowledges the support from CNPq. J.M. and D.L.O. thank the support from FAPESP (2018/04055-8, 2016/20667-8). A.I.K. acknowledge financial support from the Australian Research Council (discovery project 170100521) and from the Australian Research Council Centre of Excellence for All Sky Astrophysics in 3 Dimensions (ASTRO 3D), through project number CE170100013.

\textit{Facilities:} \textbf{ESO}: VLT 8.2-meter Unit Telescopes, Echelle SPectrograph for Rocky Exoplanet and Stable Spectroscopic Observations (ESPRESSO).

\section*{Data Availability}
The datasets were derived from sources in the public domain:  \url{http://archive.eso.org/wdb/wdb/eso/espresso/form} (under the ESO program IDs 0102.C-0523 and 1102.A-0852).





\bibliographystyle{mnras}
\bibliography{references}








\bsp	
\label{lastpage}
\end{document}